\documentclass[nohyperref]{article}

\usepackage{microtype}
\usepackage{booktabs} 
\usepackage{hyperref}

\usepackage[accepted]{icml2022}

\usepackage{amsmath}
\usepackage{amssymb}
\usepackage{mathtools}
\usepackage{amsthm}

\usepackage[capitalize,noabbrev]{cleveref}

\theoremstyle{plain}

\theoremstyle{definition}

\theoremstyle{remark}

\usepackage[textsize=tiny]{todonotes}

\usepackage{caption}
\usepackage{subcaption}
\usepackage{graphicx}

\icmltitlerunning{Synthesizing Personalized Non-speech Vocalization from Discrete Speech Representations}

\begin{document}

\twocolumn[
\icmltitle{
    Synthesizing Personalized Non-speech Vocalization from \\
    Discrete Speech Representations
}



\icmlsetsymbol{equal}{*}

\begin{icmlauthorlist}
    \icmlauthor{Chin-Cheng Hsu}{comp}
\end{icmlauthorlist}

\icmlaffiliation{comp}{Resemble AI, Toronto, Canada}

\icmlcorrespondingauthor{Chin-Cheng Hsu}{jeremy@resemble.ai}

\icmlkeywords{Machine Learning, ICML, Voice Conversion}

\vskip 0.3in
]




\printAffiliationsAndNotice{}

\begin{abstract}
We formulated non-speech vocalization (NSV) modeling as a text-to-speech task
and verified its viability.
Specifically, we evaluated the phonetic expressivity of HUBERT speech units
on NSVs 
and verified our model's ability to 
control over speaker timbre
even though the training data is speaker few-shot.
In addition, we substantiated that the heterogeneity in recording conditions is the major obstacle for NSV modeling.
Finally, we discussed five improvements over our method for future research.
Audio samples of synthesized NSVs are available
on our demo page: \url{https://resemble-ai.github.io/reLaugh}.
\end{abstract}

\section{Introduction}
Non-speech vocalizations (NSVs) 
play an important role in 
daily conversations,
voice acting, and
paralinguistic communication \cite{paraling}.
Nonetheless, due to their diversity, spontaneity, and lack of non-verbal forms,
NSV modeling has not been as successful as speech and remains a challenge to this day
\cite{DBLP:conf/icassp/UrbainCD13, LaughGANter, tiis:24633}.
One of the major challenges has been the lack of data.

In response to the needs,
the ExVo Workshop 2022 \cite{ExVo} was hosted 
to provide the research community with a dataset and benchmarks.
The ExVo dataset features 
a collection of over 1,500 speakers,
10 soft labels for emotions, and
over 36 hours of speech in total duration.
Three sub-challenges were set,
aiming at solutions to multi-task learning, 
few-shot learning, and 
synthesis (ExVo Generate).

\begin{figure}
\centering
  \begin{subfigure}[b]{0.5\textwidth}
    \centering
    \includegraphics[width=1\linewidth]{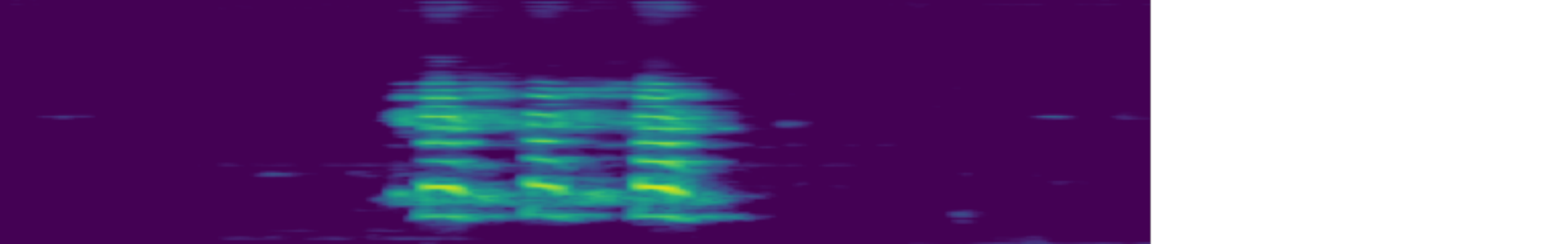}
    \caption{
        Utterance \texttt{00000} has a cut-off frequency at about 8 kHz and appears to have an excessive dynamic range compression.
    }
    \label{fig:mel-a}
  \end{subfigure}
  \begin{subfigure}[b]{0.5\textwidth}
    \centering
    \includegraphics[width=1\linewidth]{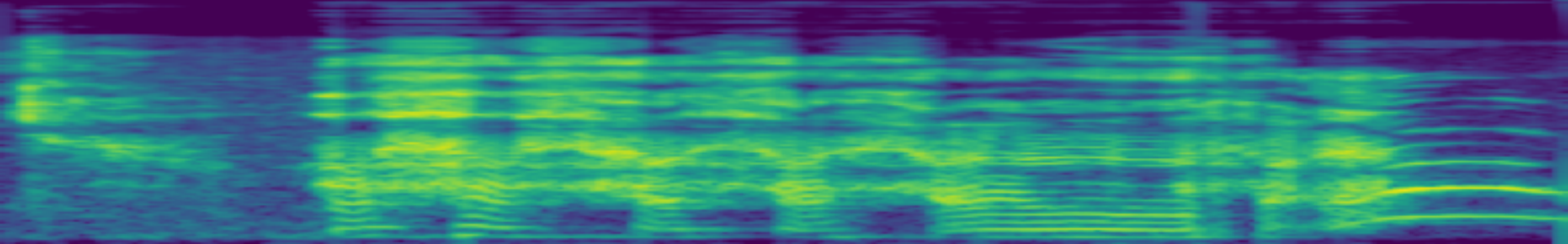}
    \caption{
        Utterance \texttt{16543} has a severe reverberation, and the cut-off frequency is around 10 kHz.
    }
    \label{fig:mel-b}
  \end{subfigure}
  \begin{subfigure}[b]{0.5\textwidth}
    \centering
    \includegraphics[width=1\linewidth]{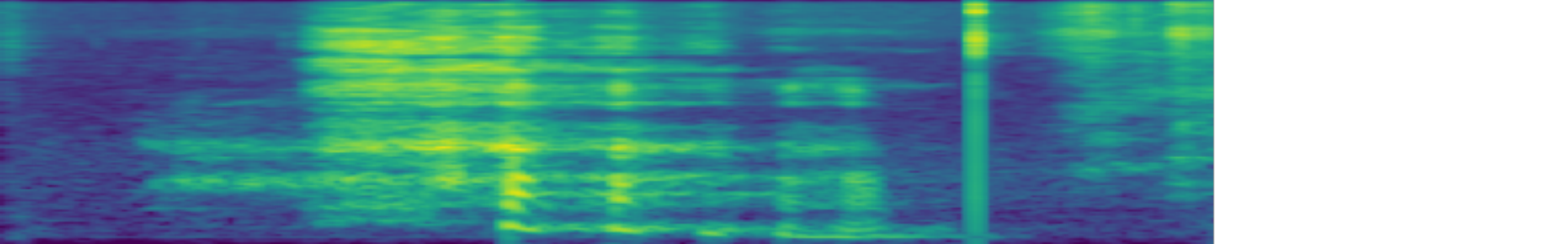}
    \caption{
        Utterance \texttt{04995} has background noise, reverberation, and a mouse click (the vertical stripe).
    }
    \label{fig:mel-c}
  \end{subfigure}
  \caption{
    Mel-spectrograms of ExVo clips with different recording conditions. All of the three clips are laughter found in \texttt{Train} set. 
  }
\label{fig:mels}
\end{figure}

Despite the large volume,
the ExVo dataset exhibits
two properties which make NSV modeling challenging.
First, each speaker has only a limited amount of data (slightly above 1 minute per speaker on average).
As a consequence,
building a model for a particular speaker become 
a difficult few-shot learning problem.
The fact that the 10 types (emotions) of vocalizations in this dataset are highly different
only makes the few-shot problem worse. 
Second, the recording conditions of these in-the-wild audio clips are heterogeneous: 
each speaker has a unique recording condition, which includes
microphone characteristics,
codec,
dynamic range compression,
sampling rates,
room reverberations,
and background noise
(see Figure \ref{fig:mels} for examples).
A diverse dataset may improve the robustness of a classifier, but it puts synthesizer training at risk in the lack of sufficient labeling.

We propose to deal with the few-shot problem by 
leveraging the shared phonetic content in NSVs.
The assumption we make is that, like speech, NSV can be decomposed into phonetic content and speaker timbre.
Based on this assumption, we tackle NSV synthesis using a
multi-speaker text-to-speech (TTS) system.
The risks are 
are two-fold: 
First, how to obtain \textit{text} for non-speech vocalizations? 
I the \textit{text} we adopt sufficiently expressive? 
Second, can a TTS learn a meaningful speaker space from speaker few-shot, in-the-wild NSV data?
Our experiments show that we can synthesize from text-like inputs 
certain types of NSV.
Moreover, our model can control over the speaker identity of synthesized NSVs.
Finally, our model learns a speaker space characterized by recording conditions, 
manifesting the difficulty induced by heterogeneity.

\section{Method}
Our method can be divided into three modules.
First, we will describe how to ``transcribe'' an audio clip into a ``text-like'' sequence using HUBERT \cite{hubert} in Sec. \ref{sec:voc-enc}.
Then we will describe in Sec. \ref{sec:am} an acoustic model predicts log-magnitude Mel-frequency spectrograms (later referred to as \texttt{mels})
from the aforementioned ``text''. 
Lastly, we will describe our vocoder that converts \texttt{mels} into waveforms in Sec. \ref{sec:voc}.

\subsection{Transcription}
\label{sec:voc-enc}
To get the ``text'' out of an NSV clip,
we resort to 
a pre-trained HUBERT model and a pre-trained K-means module with 100 clusters.
HUBERT first converts the input audio into a sequence of feature vectors, and then K-means discretizes each vector by representing it with the index of the the closest cluster centroid.
We encode the index sequence (HUBERT speech units) by run-length, splitting it into two sequences of the same lengths:
one is a sequence of speech units that are not repeating consecutively, and 
the other is the \textit{duration} sequence recording the number of repetitions. 
Finally, we map the index to a character.
As a result, we transcribed an NSV into ``text'' which is easier to read and to search from.
We refer to final ``text'' as \textit{pseudo-phonemes} because they serve the same purpose as text or phonemes in a TTS system.
The procedures are illustrated in Figure \ref{fig:arch-enc}.

\subsection{Acoustic Modeling}
\label{sec:am}
We convert pseudo-phonemes into \texttt{mels} using an architecture inspired from FastSpeech2 \cite{fastspeech2} 
which consists of an encoder, a variance adaptor, and a decoder.
We made the following modifications:
First, we replaced all feed-forward transformers with dilated convolution stacks.
Second, we kept only the duration predictor in the variance adaptor.
Third, we added a pitch decoder in parallel to the \texttt{mel} decoder; it outputs raw pitch values (scaled to $[0, 1]$) instead of wavelet transform coefficients.

\begin{figure}
\centering
  \begin{subfigure}[b]{0.25\textwidth}
    \centering
    \includegraphics[width=0.95\linewidth]{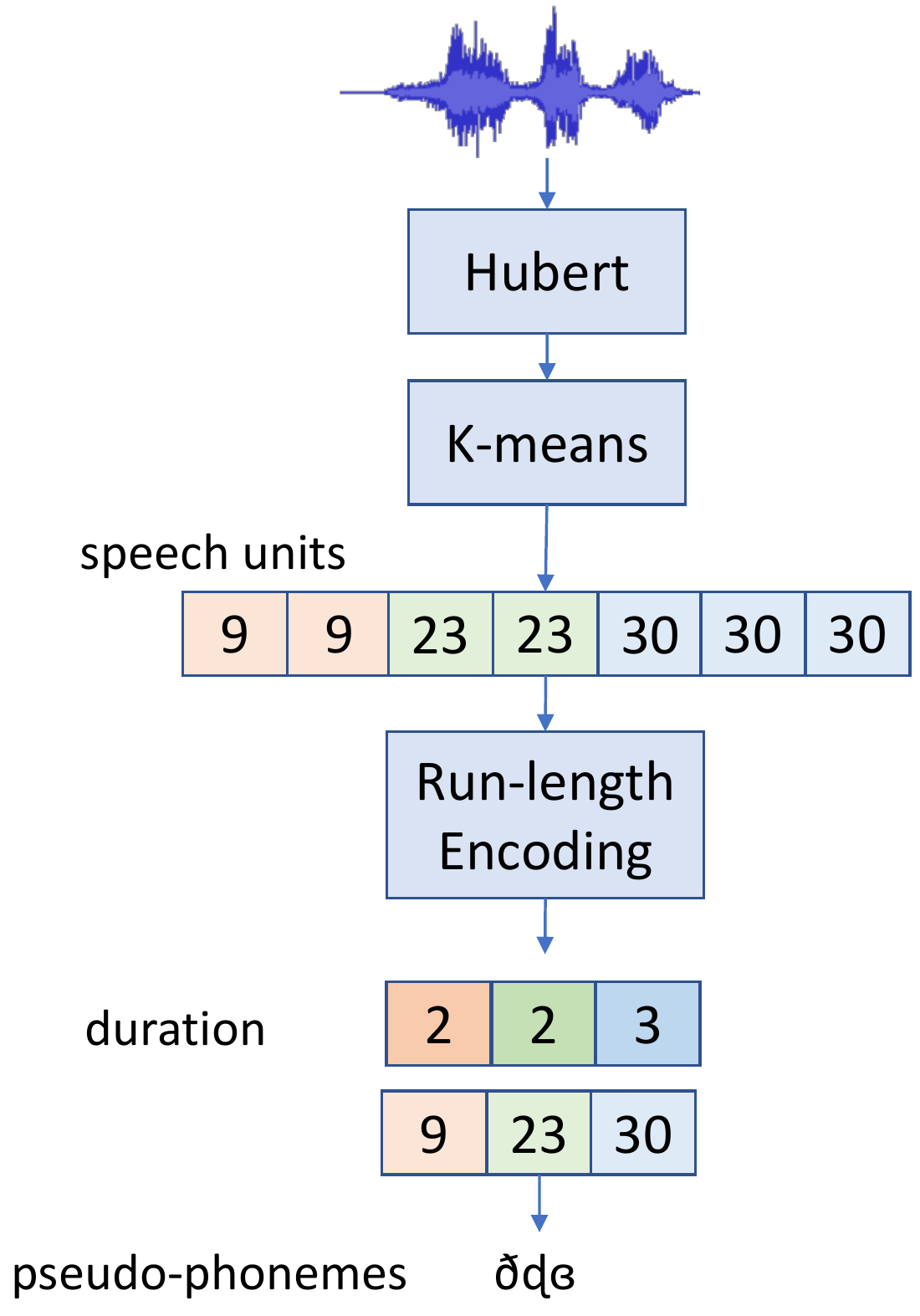}
    \caption{Transcriber.}
    \label{fig:arch-enc}
  \end{subfigure}%
  \begin{subfigure}[b]{0.25\textwidth}
    \centering
    \includegraphics[width=0.95\linewidth]{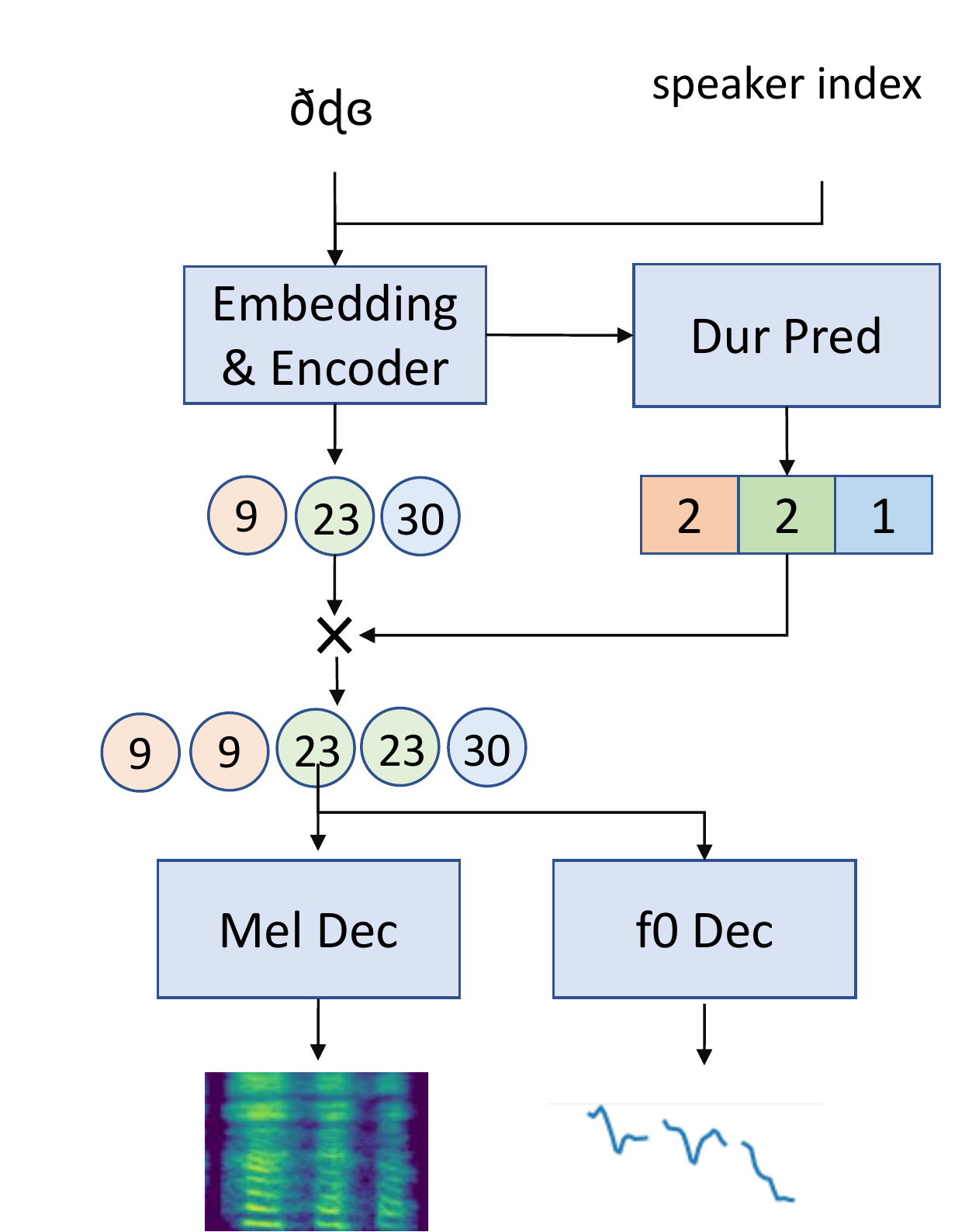}
    \caption{Acoustic model.}
    \label{fig:arch-trn}
  \end{subfigure}%
  \caption{
  Illustration of the vocalization encoding procedures and acoustic model.
  }
  \label{fig:arch}
\end{figure}

The input to our acoustic model is a pseudo-phoneme sequence and a speaker label.
During training, we up-sample the pseudo-phoneme sequence to the same length as the \texttt{mels}
using the duration sequence we extracted in Sec. \ref{sec:voc-enc}.
During synthesis, 
we up-sample the pseudo-phoneme sequence using predicted duration.
Finally, our acoustic model predicts the pitch contour and \texttt{mels} from the up-sampled pseudo-phoneme sequence.
The procedures are illustrated in Figure \ref{fig:arch-trn}.


\subsection{Vocoder}
\label{sec:voc}
We synthesize waveforms from \texttt{mels} and pitch contour using HooliGAN \cite{hooligan},
a neural vocoder based on harmonic-plus-noise model (HNM) \cite{hnm}.
The HNM treats speech signal as the superposition of periodic and aperiodic components.
HooliGAN inherits the idea of modeling the periodic signals (harmonics) using an oscillator
and thus requires the pitch contour for driving the oscillator.
The aperiodic components, on the other hand, are modeled by a convolutional network conditioned on \texttt{mels}.
We are interested in whether HooliGAN is suitable for modeling NSVs
whose periodic components are less consecutive compared to those in normal speech.

\section{Experiments}

\subsection{Dataset and Processing}
The ExVo dataset is split into three subsets:
the training set (\texttt{Train}),
the validation set (\texttt{Val}),
and the test set.
The \texttt{Train} and \texttt{Val} sets can be further divided into 10 subsets by treating the emotion class with the largest intensity as the exclusive emotion label for each audio clip.
We restricted ourselves to using only the \texttt{Train} set for training, in accordance with the ExVo white paper \cite{ExVo}.

The \texttt{Train} set covers 571 crowd-sourced speakers, each of which has a different recording environment with his or her own equipment.
The total duration of the subset is 12 hours.
We excluded silent clips and speakers whose clips all have an extremely low volume.
Clips with \texttt{Triumph} and \texttt{Horror} emotion labels were discarded due to their extreme scarcity (less than 1 and 2 clips per speaker on average, respectively).
In the end, about 6\% of the \texttt{Train} set were removed.
After pruning, we converted \texttt{webm} into \texttt{wav} files and down-sampled the them from 48 to 32 kHz.
We extracted pitch and \texttt{mel} 
using 10 ms frame rate, 
25 ms frame length,
and 256 Mel banks.
We trained both our acoustic model and HooliGAN models from scratch solely on the pruned \texttt{Train} set 
for 50,000 and 100,000 steps, respectively.

\begin{figure}
\centering
  \includegraphics[width=1\linewidth]{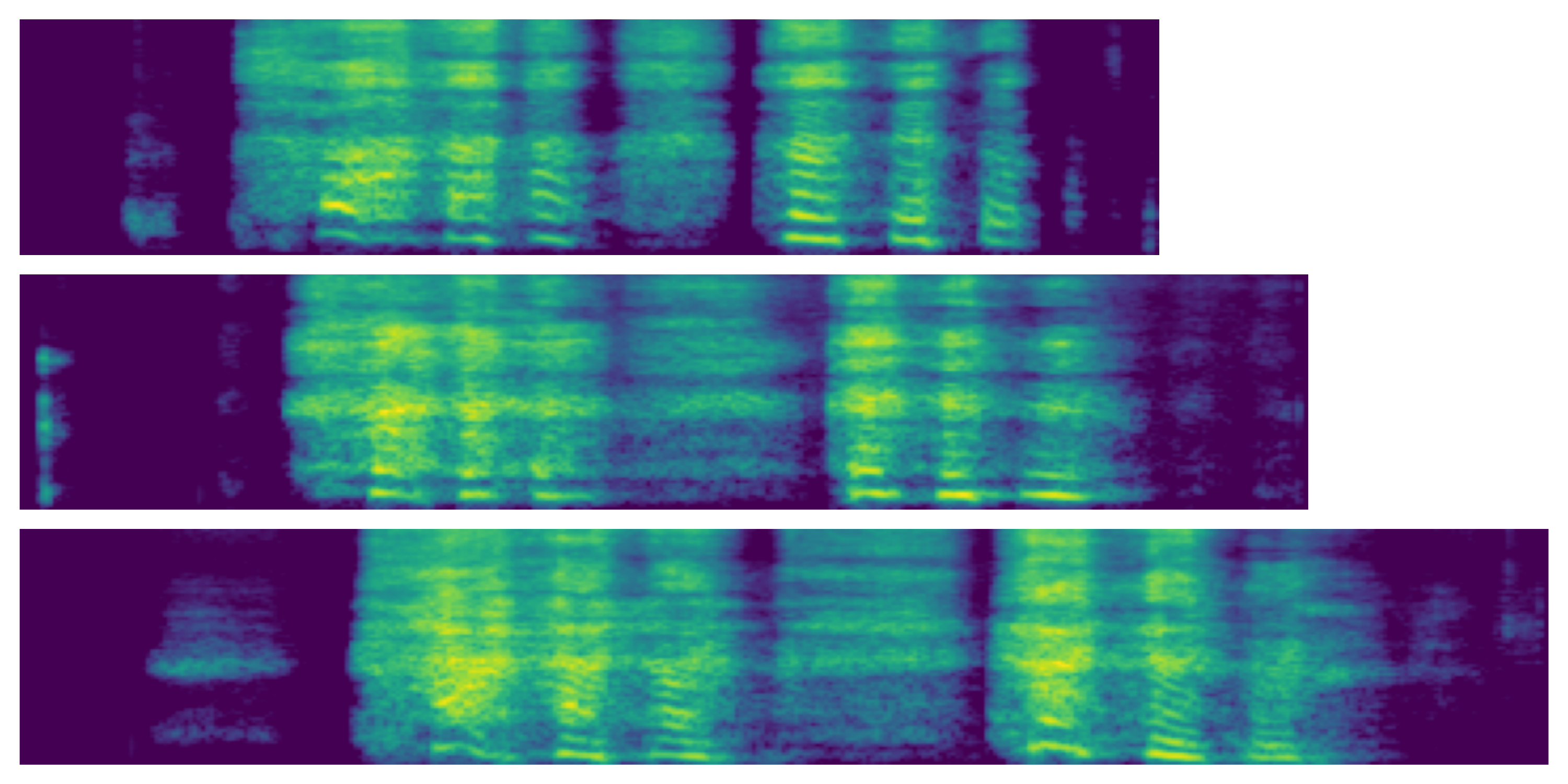}
  \caption{
  Mel-spectrograms of synthesized laughter from the same pseudo-phoneme sequence with three different speaker labels. 
  The phonetic content (hahaha [inhaling] hahaha) remains the same across syntheses.
}
\label{fig:vc}
\end{figure}

\subsection{Synthesis}
\label{sec:syn}
To generate an NSV for evaluation, 
we first sample a pseudo-phoneme sequence from the training set
and replace its speaker label with another randomly sampled speaker.
The acoustic model first predicts the speaker-dependent duration of each pseudo-phoneme and then 
up-samples the pseudo-phoneme sequence using the predicted duration.
Then, it predicts \texttt{mel} and pitch contour from the up-sampled pseudo-phoneme embeddings conditioned on the speaker label.
Finally, the vocoder predicts the waveform from the \texttt{mel} and the pitch contour.
Figure \ref{fig:vc} showcases \texttt{mels} of the synthesized samples.

\subsection{Evaluations}
We submitted 100 synthesized NSV audio files for \textit{Amusement} and \textit{Awe} classes, respectively, to ExVo Generate.
The Human-Evaluated Expression Precision (HEEP) we received from the ExVo organizer team is listed in Table \ref{tab:heep}.
Our submission scores higher in Amusement compared to the baseline, indicating that our model is better at modeling laughter.
To our surprise, our submission does not score high on Awe,
which are highly speech-like (wow, whoa, woo) both in the training data and in the syntheses.

Following \cite{ExVo}, 
we also evaluated our results using Fr{\'e}chet Inception Distance (FID) \cite{FID}
presented in Table \ref{tab:fid}\footnote{
  For internal consistency, we report FID based on the stats computed on the data we processed. 
  For the competition, we computed FID of our submission using the stats in \href{https://github.com/HumeAI/competitions/tree/main/ExVo2022}{ExVo's official repository}, yielding 1.417 (Amusement) and 2.707 (Awe). 
}.
The reference is the \texttt{Val} set, and 
we report only the emotion-specific FIDs where two input sets have the same emotion label so that the FID values are not affected by the discrepancy of phonetic content.
We also report the FID between the \texttt{Train} and the \texttt{Val} sets which can be considered the performance upper-bound.
Due to the stochasticity in our synthesis procedures, 
we repeated the valuation for 10 times and reported the mean and standard deviation.
The results have two indications:
First, FID can be highly inaccurate and biased when the sample size is small.
Second, there is still large room for improvements.

\begin{table}[t]
\caption{
  HEEP (Human-Evaluated Expression Precision $\uparrow$) of our submission to ExVo Generate.
}
\label{tab:heep}
\vskip 0.15in
\begin{center}
\begin{tabular}{l|cc}
\toprule
Emotion    &  Ours  &  \cite{ExVo}  \\
\midrule
Amusement  & 0.79  & 0.49 \\ 
Awe        & 0.51  & 0.46 \\ 
\bottomrule
\end{tabular}
\end{center}
\vskip -0.1in
\end{table}

\begin{table}[t]
\caption{
  Fr{\'e}chet inception distance (FID $\downarrow$) from  \texttt{Val} set.
  The prefix ``Syn'' denotes the synthesized outputs from our model,
  and the suffix denotes the number of utterance in the input set.}
\label{tab:fid}
\vskip 0.15in
\begin{center}
\begin{tabular}{l|ccc}
\toprule
Emotion  &  \texttt{Train}  &  Syn-1000  & Syn-100\\
\midrule
Amusement  & 0.15  & 0.52 $\pm$ 0.03  & 0.81 $\pm$ 0.11 \\ 
Awe        & 0.18  & 0.69 $\pm$ 0.06  & 1.23 $\pm$ 0.16 \\ 
Awkward    & 0.27  & 0.76 $\pm$ 0.05  & 1.19 $\pm$ 0.07 \\ 
Distress   & 0.18  & 1.00 $\pm$ 0.06  & 1.43 $\pm$ 0.12 \\ 
Excitement & 0.20  & 1.02 $\pm$ 0.04  & 1.86 $\pm$ 0.11 \\ 
Fear       & 0.12  & 0.50 $\pm$ 0.05  & 1.06 $\pm$ 0.10 \\ 
Sadness    & 0.22  & 0.44 $\pm$ 0.03  & 0.94 $\pm$ 0.09 \\ 
Surprise   & 0.08  & 0.39 $\pm$ 0.03  & 0.86 $\pm$ 0.10 \\ 
\midrule
Horror     & 0.41  & -     & - \\
Triumph    & 1.37  & -     & -\\
\bottomrule
\end{tabular}
\end{center}
\vskip -0.1in
\end{table}

\subsection{Recording Condition Characterizes the Speaker Space}
\begin{figure}
\centering
  \begin{subfigure}[b]{0.25\textwidth}
    \centering
    \includegraphics[width=0.95\linewidth]{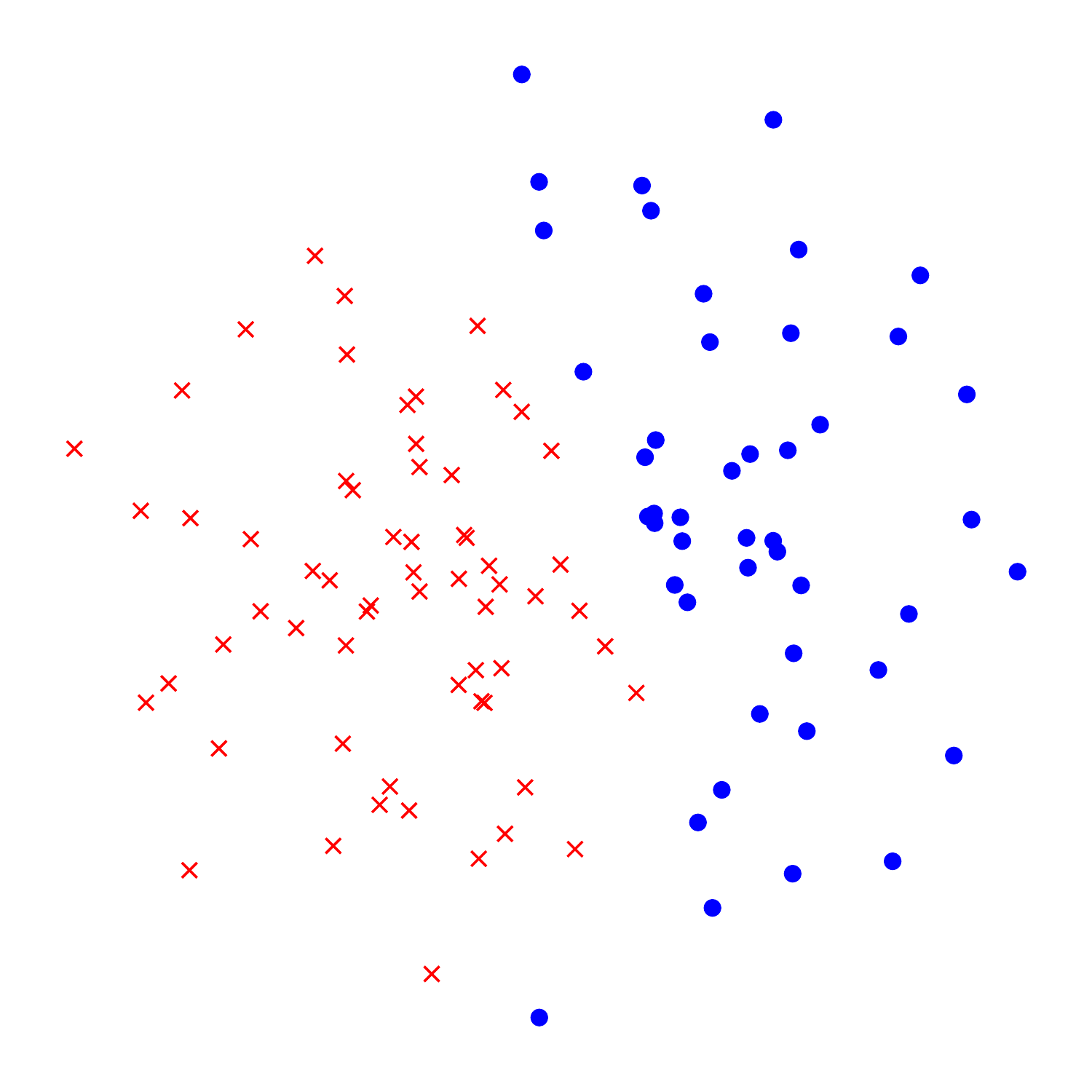}
    \caption{VCTK}
    \label{fig:spk-vctk}
  \end{subfigure}%
  \begin{subfigure}[b]{0.25\textwidth}
    \centering
    \includegraphics[width=0.95\linewidth]{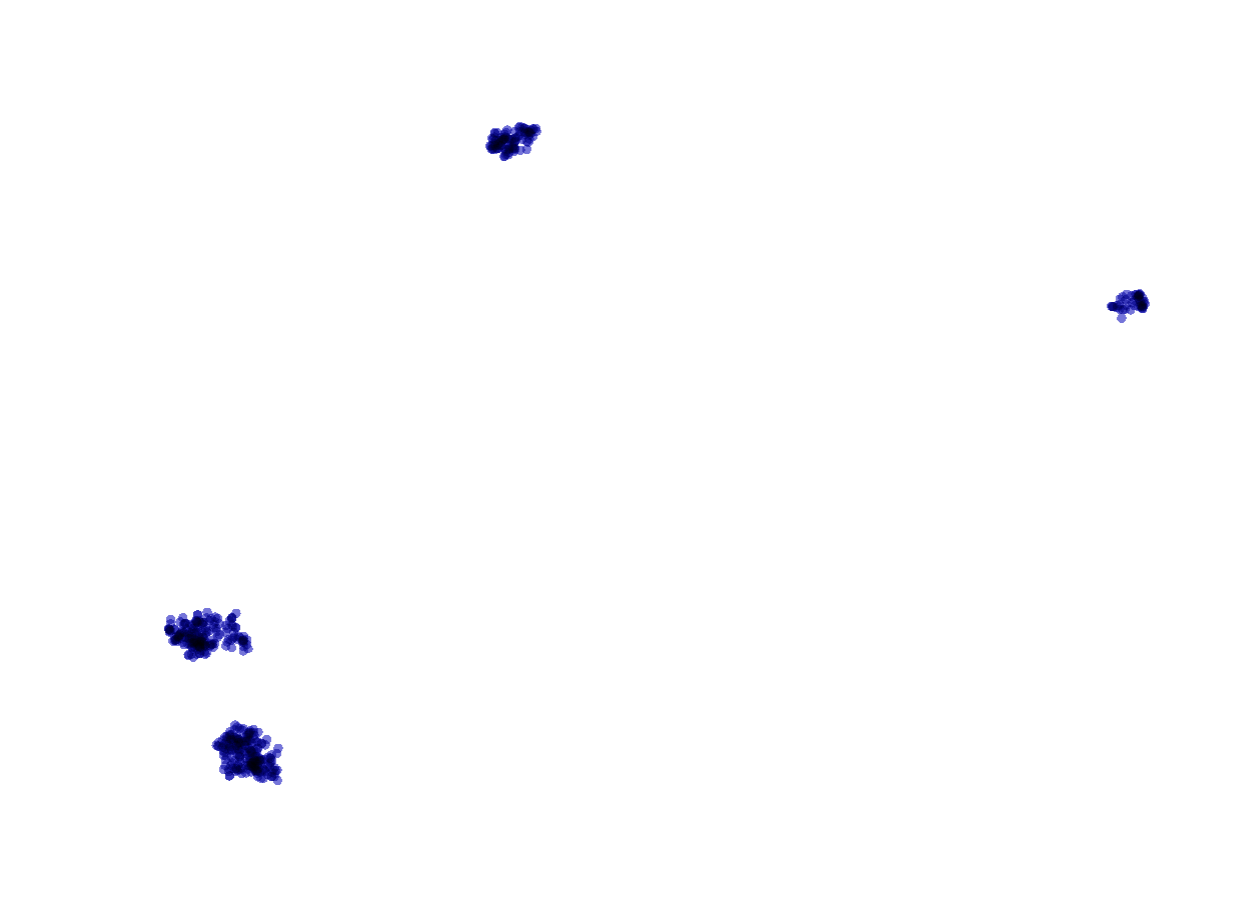}
    \caption{ExVo}
    \label{fig:spk-emb-exvo}
  \end{subfigure}%
  \caption{
  Scatter plot of speaker embeddings on a two-dimensional t-SNE space.
  ExVo speakers are clustered roughly by the recording condition of their clips 
  whereas VCTK speakers divide
  into gender groups.
  (Blue dot: male. Red cross: female).
  }
  \label{fig:spk-emb}
\end{figure}

We found recording conditions characterized the speaker space our model learned.
In our model, a speaker is represented by a randomly initialized vector and is updated by back-propagation.
Notwithstanding the simplicity, 
our model learns to embed speakers whose characteristics are similar at the same neighborhood.
In Figure \ref{fig:spk-emb}, we compared two speaker spaces our model learned.
For a standard TTS dataset such as VCTK \cite{vctk} where all the clips are recorded in a hemi-anechoic chamber, 
the speaker space is rather simple:
the speaker embeddings divide into two gender groups, possibly separable by a simple hyperplane.
What differs ExVo from a standard TTS dataset is the heterogeneous recording conditions,
which is manifested in in Figure \ref{fig:spk-emb-exvo}.

The snapshot in Figure \ref{fig:spk-emb-exvo} has 4 clusters:
In Cluster 1, most of the clips have a cut-off frequency at 8 kHz (see Figure \ref{fig:mel-a} for an example).
In Cluster 2, most of the clips have a low quality in general;
In contrast, clips in Cluster 3 have higher sampling rates, but they have various defects, such as severe reverberations or aliasing above 8 kHz.
Cluster 4 covers most of the clips that are originally Opus \cite{opus} fullband (cut-off frequency at 20 kHz). 
The embedding projector can be found on our demo page.

\subsection{Transferring Speaker Identity of an NSV}
Finally, we tested our model's ability to control over speaker timbre.
As mentioned in Sec. \ref{sec:syn}, 
we can change the timbre by specifying the desired speaker label when generating NSVs.
The question is whether the phonetic content remains the same.
By listening to syntheses\footnote{
    Due to the limitation of ExVo's EULA, we could not conduct listening tests for speaker similarity.
} 
(including those visualized in Figure \ref{fig:vc}),
we can verify that the phonetic content does remain unaltered.
The meaning of this observation is two-fold:
First, pseudo-phonemes are sufficiently expressive for  representing some NSVs
because the same pseudo-phoneme inputs result in the same phonetic content in the syntheses.
Second, our model can capture speaker timbre from NSV and can convert speaker identities.
These results shed light on voice conversion and speaker modeling from NSVs.

\subsection{Future Work}
Our method can be improved from five aspects.
First, incorporating another large set of high-quality recordings into ExVo may help the model to learn a better template and generalize from it.
Second, methods, such as \cite{gmvae}, that disentangle recording condition and speaker identity may allow us to remove the negative effects from recording conditions.
Third, pre-training the TTS and vocoder models on large datasets (speech or non-speech) may get the models a better initialization and thus better quality.
Fourth, the speech units from HUBERT might be suboptimal because HUBERT is trained on speech data only.
Finally, building a ``language model'' over pseudo-phonemes will enable the model to do free generation.

\section{Conclusions}
We demonstrated a multi-speaker TTS method for synthesizing non-speech vocalizations from HUBERT's discrete representations. 
Our experiments answered two questions about applying this method to NSV modeling.
First, we found HUBERT speech units, which were treated as an alphabet of 100 letters in this work, capable of characterizing NSVs.
Second, we found that our model learned a speaker space characterized by recording conditions.
Our results also implied the viability of speaker modeling and voice conversion from NSV data.
Finally, we proposed possible ways for improving the method for future research.

\bibliography{myref}
\bibliographystyle{icml2022}

\newpage
\appendix
\onecolumn


\end{document}